\documentclass[aps,prd,twocolumn,nofootinbib]{revtex4}
\usepackage{epsfig}
\usepackage{slashed}

\newcommand{\bel}[1]{\begin{equation}\label{#1}}
\newcommand{\bal}[1]{\begin{eqnarray}\label{#1}}

\newcommand{\be}{\begin{equation}}
\newcommand{\ee}{\end{equation}}
\newcommand{\ba}{\begin{eqnarray}}
\newcommand{\ea}{\end{eqnarray}}

\newcommand{\bes}{\begin{equation*}}
\newcommand{\ees}{\end{equation*}}

\begin{document}
%%%%%%%%%%%%%%%%%%%%%%%%%%%%%%%%%%%%%%%%%%
\title{Droplets in the cold and dense linear sigma model with quarks}

\author{Let\'\i cia F. Palhares\footnote{leticia@if.ufrj.br} and 
Eduardo S. Fraga\footnote{fraga@if.ufrj.br} }

\affiliation{Instituto de F\'\i sica, Universidade Federal do Rio de Janeiro, \\
Caixa Postal 68528, Rio de Janeiro, RJ 21941-972, Brazil}

\begin{abstract}
The linear sigma model with quarks at very low temperatures provides an effective 
description for the thermodynamics of the strong interaction in cold and dense matter, 
being especially useful at densities found in compact stars and protoneutron star matter. 
Using the $\overline{\rm MS}$ one-loop effective potential, we compute quantities that are 
relevant in the process of nucleation of droplets of quark matter in this scenario. In particular, 
we show that the model predicts a surface tension of $\Sigma \sim 5$--$15~$MeV/fm$^{2}$, 
rendering nucleation of quark matter possible during the early post-bounce stage of core collapse supernovae. 
Including temperature effects and vacuum logarithmic corrections, we find a clear competition between these
features in characterizing the dynamics of the chiral phase conversion, so that if the temperature is low enough
the consistent inclusion of vacuum corrections could help preventing the nucleation of quark matter during the collapse process.
We also discuss the first interaction corrections that come about at two-loop order.
%and consequences on the location of the critical chemical potential and other thermodynamic quantities.
\end{abstract}

\maketitle

%%%%%%%%%%%%%%%%%%%%%%%%%%%%%%%%%%%%%%%%%%
\section{Introduction}

The thermodynamics of strong interactions for cold matter under extremely high 
densities is of utmost importance for the understanding of the structure of compact 
stars \cite{stars}. Since quantum chromodynamics (QCD) is asymptotically free, it is 
believed that for high enough densities quarks will be in a deconfined state, the 
quark-gluon  plasma \cite{Rischke:2003mt}. Moreover, due to the approximately chiral 
nature of the QCD action, one also expects quarks to be essentially massless above a 
sufficiently high value of the chemical 
potential\footnote{Nevertheless, these features 
and other thermodynamic properties of dense media were also shown to be significantly 
affected by nonzero quark masses \cite{Fraga:2004gz,Kurkela:2009gj}.}. 
Depending on the location of the critical 
density, one might find several sorts of condensates and even deconfined quark matter 
in the core of neutron stars \cite{Alford:2006vz}. Furthermore, the order and strength of 
the chiral transition are crucial features in establishing the existence of a new class of 
compact stars \cite{Fraga:2001id,SchaffnerBielich:2004ch}. 

Recently, it was shown that deconfinement can happen early, during the early post-bounce 
accretion stage of a core collapse supernova event, which could result not only in a delayed 
explosion but also in a neutrino burst that could provide a signal of the presence of 
quark matter in compact stars \cite{Sagert:2008ka}. However, as was discussed in detail 
in Ref. \cite{Mintz:2009ay} (see also Ref. \cite{Bombaci:2009jt})  those possibilities depend on the actual dynamics of phase 
conversion, more specifically on the time scales that emerge. In a first-order phase transition, 
as is expected to be the case in QCD at very low temperatures, the process is guided by 
bubble nucleation (usually slow) or spinodal decomposition (``explosive'' due to the vanishing 
barrier), depending on how fast the system reaches the spinodal instability as compared to the nucleation 
rate. Nucleation in relatively high-density, cold strongly interacting matter, with chemical potential of 
the order of the temperature, can also play an important role 
in the scenario proposed in Ref. \cite{Boeckel:2009ej}, where a second (little) inflation at 
the time of the primordial quark-hadron transition could account for the dilution of an initially 
high ratio of baryon to photon numbers. A key ingredient in both scenarios is, of course, the 
surface tension, since it represents the price in energy one has to pay for the mere existence 
of a given bubble (or droplet). 

However, the surface tension and the whole process of phase conversion via nucleation 
in a cold and dense environment are not known for strong interactions. In fact, the 
mapping of this sector of the (equilibrium) phase diagram is still in its infancy \cite{Stephanov:2007fk}, 
let alone dynamical processes of phase conversion. This region of the phase diagram is not very 
amenable to first-principle approaches. On one hand, the so-called sign problem at finite chemical 
potential brings about major technical difficulties for performing Monte Carlo lattice simulations, 
so that one still does not have the high-quality guidance from lattice QCD that is currently 
available at finite temperature and zero density \cite{Laermann:2003cv}. On the other hand, 
although the perturbative series for QCD at zero temperature and finite density seems 
to be much better behaved than its counterpart at finite 
temperature \cite{Fraga:2001id,tony,andersen,Braaten:2002wi}, the values of chemical 
potential that are phenomenologically interesting in the interior of compact stars 
are already too low, of the order of $400-500~$MeV \cite{stars}. At this scale, one is 
certainly pushing perturbative QCD much beyond its limits of applicability. Besides, 
this approach is incapable of incorporating the chiral and the deconfinement 
transitions, as well as vacuum properties, unless it is merged with a complementary 
low-density treatment. Therefore, an estimate of the surface tension and the description 
of the nucleation process within an effective model for strong interactions in a cold and 
dense environment seems welcome. 

In this paper we consider the linear sigma model coupled with constituent quarks (LSMq) 
at zero or low temperature and finite quark chemical potential as a model for the 
thermodynamics of strong interactions in cold and dense matter. We compute 
the effective action for the sigma condensate 
% at two-loop order 
integrating 
over the quark fields and keeping quadratic fluctuations of the sigma field around the 
condensate. Our final thermodynamic potential incorporates all corrections from the 
medium and vacuum fluctuations in the $\overline{\rm MS}$ scheme, including logarithmic 
and scale-dependent contributions from quark and sigma bubble-diagrams. 
% The two-loop momentum integrals are computed analytically for {\it arbitrary} quark and 
% sigma masses, the final result being expressed in terms of well-known special functions. 
Renormalization is implemented in the standard fashion in the $\overline{\rm MS}$ scheme. 
Knowing the full thermodynamic potential, we study the behavior of the chiral condensate 
with chemical potential, the location and nature of the chiral transition, and the process 
of homogeneous nucleation.

The paper is organized as follows. Section II presents the LSMq and some of its features, 
and the analytic computation of the one-loop effective potential at finite 
chemical potential, including quantum logarithmic corrections in the vaccum and thermal 
effects. In Section III we show our results for the surface tension and other relevant 
quantities for the nucleation process, analyzing the competition between vacuum and thermal
modifications and discussing the consequences for the supernovae explosion scenario.
Section IV contains our conclusions and outlook.

%%%%%%%%%%%%%%%%%%%%%%%%%%%%%%%%%%%%%%%%%%
\section{Effective theory}

\subsection{General framework}

The LSMq, also known as the quark-meson 
model, is very suitable for the study of the chiral transition \cite{lee-book}. As argued in 
Ref. \cite{Pisarski:1983ms}, QCD with two flavors of massless quarks belongs to the same 
universality class as the $O(4)$ LSM, exhibiting the same qualitative behavior at criticality. 
Besides, the LSM is renormalizable \cite{Lee:1968da} and reproduces correctly the phenomenology 
of QCD at low energies, such as the spontaneous (and small explicit) breaking of chiral 
symmetry, meson masses, etc. In spite of the fact that an effective theory does not require 
renormalizability to be well posed, this attribute is highly desirable if one is interested in investigating 
the behavior of physical quantities as the energy scale is modified, which can be accomplished via 
renormalization group methods.

Since its proposal \cite{GellMann:1960np}, the LSM has been investigated in different 
contexts, from the low-energy nuclear theory of nucleon-meson interactions to ultra-relativistic 
high-energy heavy-ion collisions, and also in different varieties, e.g. including explicitly 
constituent quark degrees of freedom or not. The thermodynamical aspects of the model 
were first considered in the very early days of finite-temperature field theory \cite{Kirzhnits:1972ut}, 
and systematic approximations for the study of the chiral transition started to be implemented 
soon afterwards \cite{Baym:1977qb}, producing an extensive literature.

When used to mimic and study the chiral transition of QCD, the emphasis was generally on 
thermal effects \cite{Baym:1977qb,Bochkarev:1995gi,Petropoulos:1998gt,Roder:2003uz,
Scavenius:2000qd,Scavenius:2000bb,Bilic:1997sh,Marko:2010cd} 
rather than considering a cold and dense scenario, although chiral symmetry 
restoration at high densities was predicted quite early \cite{Lee:1974ma}. This choice was, 
of course, well justified by the stimulating experimental results coming from high-energy 
heavy-ion collisions \cite{QM}, and by the possibility to compare to numerical output 
from lattice QCD \cite{Laermann:2003cv}. Usually the gas of quarks is treated as a thermal 
bath in which the long-wavelength modes of the chiral field evolve, the latter playing the role 
of an order parameter in a Landau-Ginzburg approach. The standard procedure is then integrating 
over the fermionic degrees of freedom, using a classical approximation for the chiral field, to 
obtain a formal expression for the thermodynamic potential from which one can compute 
all the physical quantities of interest. In this case, the sigma field is approximated by the condensate, 
and the functional integral over sigma fluctuations is not performed. The fermionic determinant is 
usually calculated to one-loop order assuming a homogeneous and static background 
field \cite{FTFT-books}.

In a theory with spontaneous symmetry breaking, the presence of a condensate will 
modify the masses. In particular, in the case of the LSMq the masses of quarks and 
mesons will incorporate corrections that are functions of the chiral condensate, which 
is a medium-dependent quantity. Therefore, contributions from vacuum diagrams 
can not be subtracted away as trivial zero-point energies since they contain 
medium-dependent pieces via effective masses. So, although the presence of 
the medium brings no new ultraviolet divergence, in principle one has to incorporate 
carefully finite vacuum contributions that survive the renormalization procedure. These 
contributions have been very often discarded in studies of the LSMq, but were shown 
to play an important role by the authors of Ref. \cite{Mocsy:2004ab} who 
incorporate scale effects phenomenologically. Vacuum contributions were considered 
at finite density in the perturbative massive Yukawa model with analytic exact results up to 
two loops in Ref. \cite{Palhares:2008yq,thesis} and, more specifically, in optimized perturbation 
theory at finite temperature and chemical potential in Ref. \cite{Fraga:2009pi}, 
also comparing to mean-field theory. More recently, this issue was discussed in a comparison 
with the Nambu-Jona--Lasinio model \cite{Boomsma:2009eh},
in the Polyakov-LSMq model in the presence of an external magnetic field \cite{Mizher:2010zb}, 
and in the quark-meson model, with special attention to the chiral limit \cite{Skokov:2010sf}. 
%In this section, we ignore vacuum corrections, coming back to their effects afterwards.

To describe the chiral phase structure of strong interactions at finite density and vanishing 
temperature we adopt the LSMq, defined by the following lagrangian
\begin{eqnarray}
{\cal L} &=&
 \overline{\psi}_f \left[i\gamma ^{\mu}\partial _{\mu} - m_f - g(\sigma +i\gamma _{5}
 \vec{\tau} \cdot \vec{\pi} )\right]\psi_f \nonumber\\
&+& \frac{1}{2}(\partial _{\mu}\sigma \partial ^{\mu}\sigma + \partial _{\mu}
\vec{\pi}  \cdot \partial ^{\mu}\vec{\pi} )
- U(\sigma ,\vec{\pi})\;,
\label{lagrangian}
\end{eqnarray}
where
\begin{equation} 
U(\sigma ,\vec{\pi})=\frac{\lambda}{4}(\sigma^{2}+\vec{\pi}^{2} -
{\it v}^2)^2-h\sigma
\label{bare_potential}
\end{equation}
is the self-interaction potential for the mesons, exhibiting both spontaneous 
and explicit breaking of chiral symmetry. The $N_f=2$ massive fermion fields 
$\psi_f$ represent the up and down constituent-quark fields $\psi_{f}=(u,d)$. For 
simplicity, we attribute the same mass, $m_f=m_q$, to both quarks. The scalar field 
$\sigma$ plays the role of an approximate order parameter for the chiral transition, 
being an exact order parameter for massless quarks and pions. The latter are 
represented by the pseudoscalar field $\vec{\pi}=(\pi_{1},\pi_{2},\pi_{3})$. It is 
customary to group together these meson fields into a $O(4)$ chiral field 
$\phi =(\sigma,\vec{\pi})$. For simplicity, we discard the pion dynamics from 
the outset, knowing that they do not affect appreciably the phase conversion 
process \cite{Scavenius:2000bb}, and focus our discussion in the quark-sigma sector 
of the theory. Nevertheless, pion vacuum properties will be needed to fix the 
parameters of the lagrangian later. 
As will be discussed below, the parameters of the lagrangian are chosen such that 
the effective model reproduces correctly the phenomenology of QCD at low energies 
and in the vacuum, such as the spontaneous (and small explicit) breaking of chiral 
symmetry and experimentally measured meson masses. 

Since chiral symmetry is spontaneously broken in the vacuum, it is convenient 
to expand the $\sigma$ field around the condensate, writing 
$\sigma(\vec{x},\tau) =\langle\sigma\rangle + \xi (\vec{x},\tau)$, where 
$\tau$ is the imaginary time in the Matsubara finite-temperature formalism. 
Although we assume the system to be at temperature $T=0$, it is convenient to 
compute the loop expansion at finite temperature, taking the limit 
$T\rightarrow 0$ at the end. For a dense system, the chiral condensate, 
$\langle\sigma\rangle$, will be a function of the quark chemical potential, 
$\mu$. Given the shift above, the fluctuation field $\xi$ is such that 
$\langle\xi\rangle =0$ and $\xi (\vec{p}=0,\omega=0)=0$. From the 
phenomenology, one expects that 
$\langle\sigma\rangle (\mu \rightarrow \infty) \approx 0$ (being equal in 
the case of massless quarks).

Keeping terms up to $O(\xi^2)$, one obtains the following effective 
lagrangian
\begin{eqnarray}
{\cal L}  &=&
\overline{\psi}_f \left[i\gamma ^{\mu}\partial _{\mu} - 
M_q - g\xi \right]\psi_f \nonumber \\
&+& \frac{1}{2}\partial _{\mu}\xi \partial ^{\mu}\xi 
-\frac{1}{2}M_\sigma^2\xi^2 - U(\langle\sigma\rangle)\;,
\label{effective_lagrangian}
\end{eqnarray}
where we have defined the $\mu$-dependent effective masses
\begin{equation}
M_q \equiv m_q + g \langle\sigma\rangle \;\;,\;\; 
M_\sigma^2\equiv 3\langle\sigma\rangle^2 -\lambda {\it v}^2 \; .
\label{masses}
\end{equation}
Linear terms in $\xi$ can be dropped in the action because of the 
condition $\xi (\vec{p}=0,\omega=0)=0$. 

In the effective lagrangian, the medium-dependent masses will entangle 
vacuum and medium contributions in the loop expansion, rendering the 
renormalization process more subtle. This is a consequence of the presence 
of a nonzero $\mu$-dependent condensate in the broken phase. Of course, the 
ultraviolet divergences are the ones of the theory in the vacuum, and 
renormalization is implemented in the standard fashion by adding 
medium-independent counterterms to the original lagrangian (\ref{lagrangian}). 
In our case, though, it is more convenient to introduce counterterms in 
the effective theory, so that there are contributions from pure-vacuum, 
pure-medium and vacuum-medium pieces in the cancellation of ultraviolet 
divergences. Renormalization is then implemented by using standard methods 
of finite-temperature field theory \cite{FTFT-books}.

%%%%%%%%%%%%%%%%%%%%%%
\subsection{Cold and dense 1-loop effective potential and vacuum corrections}

The effective potential can be computed exactly and in closed form following 
the steps detailed in Ref. \cite{Palhares:2008yq}, a procedure that can be even 
generalized to optimized perturbation theory \cite{Fraga:2009pi}. Keeping only 
contributions to one loop order, the effective potential is given by
\begin{equation}
V_{\rm eff}(\bar\sigma)=U(\bar\sigma)+\Omega^{\rm ren}_{\xi}(\bar\sigma) \,\label{Veff0}
\end{equation}
where $\Omega^{\rm ren}_{\xi}(\bar\sigma)$ is the fully-renormalized thermodynamic 
potential for the fluctuation effective theory. The latter corresponds to a Yukawa theory 
for massive fermions and a massive scalar, with masses given by (\ref{masses}) as 
functions of $\bar\sigma$. The thermodynamics of this theory was fully solved analytically 
to two loops in the cold and dense regime in Ref. \cite{Palhares:2008yq}, where all details 
can be found. In the $\overline{\rm MS}$ scheme, $\Omega^{\rm ren}_{\xi}(\bar\sigma)$ can be 
written as a sum of a medium contribution
\begin{eqnarray}
\Omega_{{\rm med}}^{(1)} &=&
- N_f N_c~\frac{1}{24\pi^2}
\left\{
2\mu p_f^3 \right. \nonumber\\
&-& \left.3 M_{q}^2~\left[ \mu p_f-M_{q}^2\log\left( \frac{\mu+p_f}{M_{q}} \right) \right]
\right\}
 \label{Cold-OmegaMed1}
\end{eqnarray}
and a vacuum contribution
\begin{eqnarray}
\Omega_{{\rm vac}}^{(1)} &=&-\frac{M_{\sigma}^4}{64\pi^2}
\left[ \frac{3}{2}+\log\left( \frac{\Lambda^2}{M_{\sigma}^2} \right) \right] \nonumber\\
&+&  N_f N_c~\frac{M_{q}^4}{64\pi^2}
\left[ \frac{3}{2}+\log\left( \frac{\Lambda^2}{M_{q}^2} \right) \right]
\, ,\label{OmVac1Res}
\end{eqnarray}
where $u=\mu p_f-M_{q}^2\log\left( \frac{\mu+p_f}{M_{q}} \right)$, $p_{f}$ is the Fermi momentum 
and $\Lambda$ is the $\overline{\rm MS}$ scale. The latter can also be fixed by vacuum 
properties, as will be discussed in subsection \ref{ParFix}. 

%%%%%%%%%%%%%%%%%%%%%%%%%%%%%%%%%%%%%%%%%%
\subsection{Thermal effects}

We can also incorporate thermal effects in the calculation of the effective potential within the LSMq. 
The inclusion of the temperature dependence allows for testing the validity of the cold ($T=0$) 
approximation at low temperatures and to investigate if the thermal corrections to
nucleation parameters can play an important role in the phase conversion process.

At one loop, the well-known temperature- and density-dependent correction to the thermodynamic 
potential is that of an ideal gas of massive sigma particles and constituent quarks 
($\omega_{\sigma}^2\equiv {\bf k}^2+M_{\sigma}^2$ and $E_q^2\equiv{\bf p}^2+M_q^2$):
\begin{eqnarray}
\Omega_{{\rm med,Th}}^{(1)} &=&
T~\int\frac{d^3{\bf k}}{(2\pi)^3}
~\log\left[
1-{\rm e}^{-\omega_{\sigma}/T}
\right]
-\nonumber\\
&&
-2TN_f N_c\int\frac{d^3{\bf p}}{(2\pi)^3}
\left\{\log\left[
1+{\rm e}^{-(E_q-\mu)/T}
\right]+
\right.
\nonumber\\
&& 
\left.
+
\log\left[
1+{\rm e}^{-(E_q+\mu)/T}
\right]
\right\}
\, . \label{TH-OmegaMed1}
\end{eqnarray}
%

%%%%%%%%%%%%%%%%%%%%%%%%%%%%%%%%%%%%%%%%%%
\subsection{Parameter fixing\label{ParFix}}

As stated above, the LSMq is adopted as an effective model for QCD at low energies, so that
the parameters
%$m_q$, 
$g$, $\lambda$, $m_q$, ${\it v}$, $h$ and $\Lambda$
are fixed in order to reproduce QCD properties either measured in the vacuum or calculated numerically
via lattice QCD. Therefore, the conditions for fixing the parameters are imposed on the
vacuum effective potential. Since we aim at comparing results from cases with different vacuum effective potentials (namely
$U$ and $U+\Omega_{\rm vac}^{(1)}$), and even with different parameter sets\footnote{The case with vacuum logarithmic terms has an extra parameter: the $\overline{\textrm{MS}}$ scale $\Lambda$.}, it is useful to make explicit the parameter fixing procedure, consistently.

The conditions for fixing the model parameters are the following:

\begin{itemize}

\item The chiral condensate in the vacuum is the pion decay constant, $f_{\pi}=93~$MeV, or, in terms
of the minimum of the vacuum effective potential,
\begin{equation}
\left.\frac{\partial V_{\rm{eff}}^{\rm vac}}{\partial \langle \sigma \rangle}\right|_{
\langle \sigma \rangle=f_{\pi}} =0 \, ; \label{ddsigma}
\end{equation}

\item The partial conservation of the axial current yields
\begin{equation}
h = f_{\pi}~m_{\pi}^2 = (93 ~\textrm{MeV})~(138 ~\textrm{MeV})^2 \; , \label{h}
\end{equation}
where $m_{\pi}$ is the pion mass;

\item The current quark masses calculated in lattice QCD (cf. e.g. Ref. \cite{Chiu:2003iw}) provide:
\begin{equation}
m_q = 4.1 ~\textrm{MeV}  \; , \label{mq}
\end{equation}
which we neglect\footnote{The current quark mass also does not bring any extra qualitative feature to the model nor changes significantly the quantitative results concerning the chiral phase transition \cite{thesis}.},
setting $m_q=0$, for the sake of comparison with previous model calculations;

\item Using the constituent quark mass in the vacuum as $1/3$ of the nucleon mass 
($m_{N}=938 ~$MeV), we can fix the Yukawa coupling, $g$:
\begin{eqnarray}
&& M_q\left(\langle \sigma \rangle(p_f=0)\right)= M_q^{\rm vac} \Rightarrow \nonumber\\
&& g = \frac{1}{f_{\pi}}~\left( \frac{1}{3}~m_{N}-m_q \right) =
 3.32
 \; ; \label{g}
\end{eqnarray}

\item The value of the dressed mass of the $\sigma$ field is given by the experimental value 
of the mass of the $\sigma$ meson:
\begin{equation}
\left.\frac{\partial^2 V_{\rm{eff}}^{\rm{vac}}}{\partial \langle \sigma \rangle^2}\right|_{
\langle \sigma \rangle=f_{\pi}} = \left(M^{\rm vac}_{\sigma}\right)^2 \approx (600~\textrm{MeV})^2
 \; ; \label{d2dsigma2}
\end{equation}

\item The quark condensate is fixed by the lattice result 
(including only quarks {\it up} and {\it down}, we have \cite{Chiu:2003iw}:
$\langle \overline{\psi}_{f}\psi_{f} \rangle_{\rm vac} = -2 ~(225 ~\textrm{MeV})^3$),
so that
\begin{equation}
\left.\frac{\partial V_{\rm{eff}}^{\rm{vac}}}{\partial m_q}\right|_{{\rm vac} ; ~
\langle \sigma \rangle=f_{\pi}} = \langle \overline{\psi}_{f}\psi_{f} \rangle_{\rm vac}
\; . \label{ddmq}
\end{equation}
\end{itemize}

The conditions (\ref{h})--(\ref{g}) above fix directly the parameters $h$, $m_q$ and $g$,
independently of the inclusion of quantum corrections to the vacuum thermodynamic potential.
On the other hand, the Eqs. (\ref{ddsigma}), (\ref{d2dsigma2}) and (\ref{ddmq}) are coupled equations 
to determine the parameters ${\it v}$ and $\lambda$ (and $\Lambda$, if quantum corrections 
to the vacuum are considered) and depend on the explicit form of the vacuum effective potential. 
In the case without quantum corrections, i.e. with the vacuum effective potential being 
purely the classical potential ($V_{\rm eff}^{\rm vac}=U$), we find $\lambda^2\approx 20$ and 
${\it v}^2\approx 7696.8~$MeV$^2$. With the addition of the 1-loop vacuum 
term $\Omega_{\rm vac}^{(1)}$ to the vacuum effective potential, the solution of Eqs. (\ref{ddsigma}),
(\ref{d2dsigma2}) and (\ref{ddmq}) yields 
$\lambda^2\approx 16.65$, ${\it v}^2\approx 3296.89~$MeV$^2$ and $\Lambda \approx 16.48~$MeV.

%%%%%%%%%%%%%%%%%%%%%%%%%%%%%%%%%%%%%%%%%%
\subsection{Influence of interactions: higher-loop corrections}\label{Int}

So far, we have only included the effects of interactions indirectly in the construction of the effective model itself, 
through dressed masses and the presence of a nonzero quark condensate. However, the interaction of the sigma meson with 
the medium constituent quarks could in principle alter the predicted dynamics for the chiral transition. The incorporation 
of such corrections in the calculation of the effective potential can be implemented via the perturbative technique order by order.

The first interaction correction in the present case appears at the two-loop order. The contribution to the effective potential corresponds then to that of the thermodynamic potential of an interacting Yukawa theory with massive scalars and massive fermions, which was obtained and analyzed in detail in Ref. \cite{Fraga:2009pi}, including the vacuum contribution and nonperturbative effects within the optimized perturbation theory framework.
Therefore, one has in principle all the machinery to investigate the influence of interactions on the phase diagram of the LSMq and
the associated process of homogeneous nucleation. It should be noted that the full case, including quantum corrections in the vacuum 
effective potential, requires the (nontrivial) solution of the coupled equations (\ref{ddsigma}), (\ref{d2dsigma2}) and (\ref{ddmq})
with the 2-loop result plugged in. Being a more technical analysis, it is out of the scope of this work, concentrated on quantifying
approximately the nucleation predicted within an effective model for strong interactions at low energies. We postpone it for a future publication.
% \cite{future2loop}.

%%%%%%%%%%%%%%%%%%%%%%%%%%%%%%%%%%%%%%%%%%
\section{Surface tension and nucleation}

Let us now consider the formation of droplets of quark matter at high density 
that happens via homogeneous nucleation \cite{reviews}. Dynamically this process 
will occur either via thermal activation of droplets (thermal nucleation) or quantum 
nucleation. The physical setting we have in mind is the one that gives the best chances 
for thermal nucleation of quark droplets in cold hadronic matter found in ``hot'' protoneutron 
stars. As discussed in Ref. \cite{Mintz:2009ay}, and previously 
in Refs. \cite{Horvath:1992wq,Olesen:1993ek}, that corresponds to temperatures of the order 
of $10-20~$MeV \cite{refs-temp}. At these temperatures, and in the presence of a barrier in the effective 
potential, thermal nucleation dominates over quantum nucleation. As soon as the barrier 
disappears, the spinodal instability is reached and the mechanism that takes over is the explosive 
spinodal decomposition. The range of temperatures under consideration is, then, high enough 
to allow for thermal nucleation (quantum nucleation being comparatively 
negligible \cite{Mintz:2009ay,Bombaci:2009jt,Iida:1997ay}) and low enough to justify 
the use of the zero-temperature effective 
potential computed previously. Temperatures of a couple of tens of MeV will not modify appreciably 
the equation of state, bringing corrections $O(T^{2}/\mu^{2})\sim 1\%$ for the typical values of 
chemical potential for the system under consideration. Nevertheless, as will be shown later,
thermal corrections can be important for the process of nucleation.

%%%%%%%%%%%%%%%%%%%%%%%%%%%%%%%%%%%%%%%%%%
\subsection{Method for extracting nucleation parameters from the effective potential}

Since our framework is an effective model, we can only aim for reasonable estimates and 
functional behavior, not numerical precision. Therefore, it is convenient to work with approximate 
analytic relations by fitting the relevant region of the effective potential by a quartic polynomial 
and working in the thin-wall limit approximation for bubble nucleation. Following 
Refs. \cite{Scavenius:2000bb,Taketani:2006zg}, we can express the effective potential over the range 
between the critical chemical potential, $\mu_{c}$, and the spinodal, $\mu_{sp}$, in the familiar 
Landau-Ginzburg form
\begin{equation}
V_{\rm eff} \approx \sum_{n=0}^4 a_n \, \phi^n \;.
\label{LandGinz}
\end{equation}
Although this approximation is obviously incapable of reproducing all three 
minima of $V_{\rm eff}$, this polynomial form is found to provide a good quantitative 
description of this function in the region of interest for nucleation, i.e. where 
the minima for the symmetric and broken phases, as well as the barrier 
between them, are located. 

A quartic potential such as Eq. (\ref{LandGinz}) can always be rewritten in the form
\begin{equation}
{\mathcal V}(\varphi)=\alpha ~ (\varphi^2-a^2)^2+j\varphi \;,
\label{www}
\end{equation}
with the coefficients above defined as follows:
\begin{eqnarray}
\alpha &=& a_4\quad, \\
a^2 &=& \frac{1}{2}\left[ -\frac{a_2}{a_4}
+\frac{3}{8}\left(\frac{a_3}{a_4}\right)^2  \right]\;, \\
j &=& a_4\left[ \frac{a_1}{a_4}-
\frac{1}{2}\frac{a_2}{a_4}\frac{a_3}{a_4}+
\frac{1}{8}\left(\frac{a_3}{a_4}\right)^3   \right]\;, \\
\varphi&=&\phi + \frac{1}{4}\frac{a_3}{a_4} \; . 
\end{eqnarray}
The new potential ${\mathcal V}(\varphi)$ reproduces the original $V_{\rm eff} ( \phi )$ 
up to a shift in the zero of energy.  We are interested in the effective 
potential only between $\mu_c$ and $\mu_{sp}$.  At $\mu_c$, we will have two 
distinct minima of equal depth.  This clearly corresponds to the choice $j = 0$
in Eq.\,(\ref{www}) so that ${\mathcal V}$ has minima at $\varphi = \pm a$ and a maximum 
at $\varphi = 0$.  The minimum at $\varphi = -a$ and the maximum move closer 
together as the chemical potential is shifted and merge at $\mu_{sp}$.  Thus, 
the spinodal requires $j/\alpha a^3 = -8/3\sqrt{3}$ in Eq.\,(\ref{www}). 
The parameter $j/\alpha a^3$ falls roughly 
linearly from $0$, at $\mu=\mu_c$, to $-8/3\sqrt{3}$ at the spinodal.

The explicit form of the critical bubble in the thin-wall limit is then given 
by~\cite{Fraga:1994xd}
\begin{equation}
\varphi_b (r;\xi,R_c)=\varphi_f + \frac{1}{\xi\sqrt{2\alpha}}
\left[ 1-\tanh \left( \frac{r-R_c}{\xi} \right) \right] \;,
\label{bubproftw}
\end{equation}
where $\varphi_f$ is the new false vacuum, $R_c$ is the radius of the 
critical bubble, and $\xi=2/m$, with $m^2\equiv {\mathcal V}''(\varphi_f)$, is a measure 
of the wall thickness.  The thin-wall limit corresponds to $\xi/R_c\ll 1$ 
\cite{Fraga:1994xd}, which can be rewritten as $(3|j|/8\alpha a^3)\ll 1$.  
Nevertheless, it was shown in \cite{Scavenius:2000bb,Bessa:2008nw} for 
the case of zero density and finite temperature that the thin-wall limit becomes 
very imprecise as one approaches the spinodal (This is actually a very general 
feature of this description \cite{reviews}).In this vein, also as remarked above, 
the analysis presented below is to be regarded as semi-quantitative, and we aim 
for estimates. 

In terms of the parameters $\alpha$, $a$, and $j$ defined above, we 
find \cite{Scavenius:2000bb,Taketani:2006zg}
\begin{eqnarray}
\varphi_{t,f} &\approx& \pm a - \frac{j}{8\alpha a^2}  \quad, \\
\xi &=& \left[ \frac{1}{\alpha (3\varphi_f^2-a^2)} \right]^{1/2}
\label{twcorlength}
\end{eqnarray}
in the thin-wall limit. The surface tension, $\Sigma$, is given by
\begin{equation}
\Sigma\equiv \int_0^{\infty}{\rm d}r~\left( \frac{{\rm d}
\varphi_b}{{\rm d}r} \right)^2 
\approx \frac{2}{3\alpha\xi^3} \ ,
\end{equation}
and the critical radius is obtained from $R_c = (2\Sigma/\Delta V)$, where
$\Delta V \equiv V(\phi_f)-V(\phi_t) \approx 2 a | j |$. Finally, the free energy of a 
critical bubble is given by $F_b=(4\pi\Sigma/3)R_c^2$, and 
from knowledge of $F_b$ one can evaluate the nucleation rate 
$\Gamma \sim e^{-F_b/T}$. In calculating thin-wall properties, we shall 
use the approximate forms for $\phi_t$, $\phi_f$, $\Sigma$, and $\Delta V$ 
for all values of the potential parameters.

%%%%%%%%%%%%%%%%%%%%%%%%%%%%%%%%%%%%%%%%%%
\subsection{Results for nucleation in the cold and dense LSM}

In what follows, we consider the LSM with quarks in the absence of vacuum corrections.
The effective potential up to 1-loop order is then given by Eq. (\ref{Veff0}) with 
$\Omega_{\xi}^{\rm ren}=\Omega_{\rm med}^{(1)}$. This corresponds to the standard case, 
adopted frequently in the literature (cf. e.g. Ref. \cite{Scavenius:2000qd}).

Using the method described in the previous section, we characterize quantitatively
the nucleation process predicted within this case, calculating different nucleation
parameters. In this section, we present results for both metastable regions, above and below 
the critical chemical potential $\mu_{\rm crit}=305.03~$MeV, which correspond respectively to 
the nucleation of quark droplets in a hadronic environment and the formation of hadronic bubbles 
in a partonic medium.

\vspace{0.3cm}
%%%%%%%%%%%%%%%%%%%%%%%%%%%%%
\begin{figure}[!hbt]
\begin{center}
\vskip 0.2 cm
\includegraphics[width=8cm,angle=0]{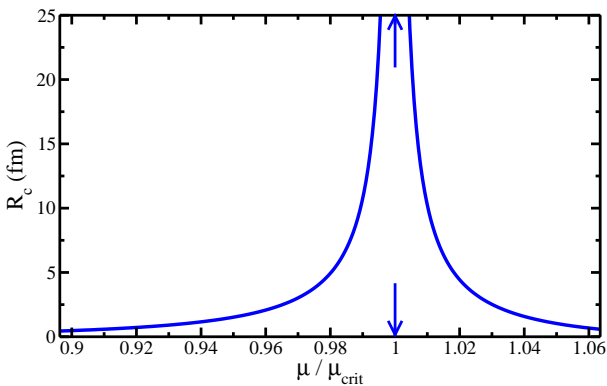}
\end{center}
\caption{Critical radius as a function of the quark chemical potential in the two metastable regions (between the spinodals).
The arrows indicate the critical chemical potential.}
\label{Rc-fig}
\end{figure}
%%%%%%%%%%%%%%%%%%%%%%%%%%%%%

In Fig. \ref{Rc-fig},
the critical radius, namely the radius of the critical bubble, is displayed. 
Any bubble created in the system via external influences or thermal 
fluctuations will either grow or shrink unless
its radius equals the critical value, which settles the threshold between suppressed and favored bubbles.
For radii bigger than the critical radius, the minimization of energy implies that the droplet will grow
and eventually complete the phase conversion. The critical radius goes to zero at the spinodal curves, where
the metastable false vacuum becomes unstable (the barrier disappears) and the phase conversion occurs explosively via the spinodal decomposition 
process\footnote{In the thin-wall approximation, which is a poor description of the regions close to the 
spinodals, the critical radius does not vanish, only becomes very small. The same is true for other 
quantities, as the surface tension.}. 
On the critical line the vacua become degenerate so that both of them are stable
yielding no nucleation, or equivalently, a divergent critical radius. In the case of the chiral transition
in the LSM with quarks, we obtain that the critical radius is $R_c<10~$fm over about $\sim 90\%$ of the metastable
regions.

%%%%%%%%%%%%%%%%%%%%%%%%%%%%%
\begin{figure}[!hbt]
\begin{center}
\vskip 0.2 cm
\includegraphics[width=8cm,angle=0]{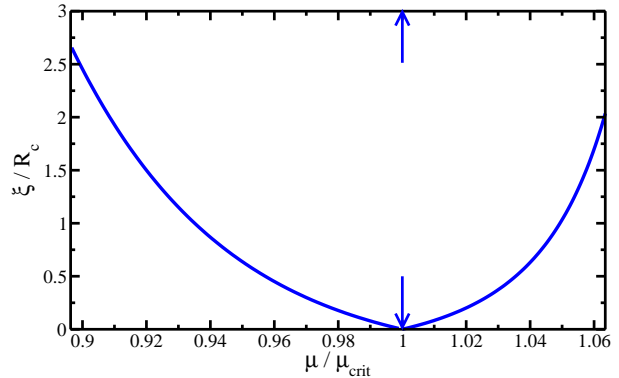}
\end{center}
\caption{Ratio between the correlation length $\xi$ and the critical radius $R_c$ as a function of the quark chemical potential in the two metastable regions.}
\label{Ratio-fig}
\end{figure}
%%%%%%%%%%%%%%%%%%%%%%%%%%%%%

The correlation length $\xi$ which provides a measure of the size of the bubble wall is plotted
in Fig. \ref{Ratio-fig}, in units of the critical radius.
As discussed in the previous section, the thin-wall approximation relies on the assumption that the 
ratio between the surface correlation length $\xi$ and the radius of the critical bubble is small.
Fig. \ref{Ratio-fig}
shows clearly that this is a reasonable condition in the vicinity of the critical
line, away from the spinodals.

\begin{figure}[!hbt]
\begin{center}
\vskip 0.2 cm
\includegraphics[width=8cm,angle=0]{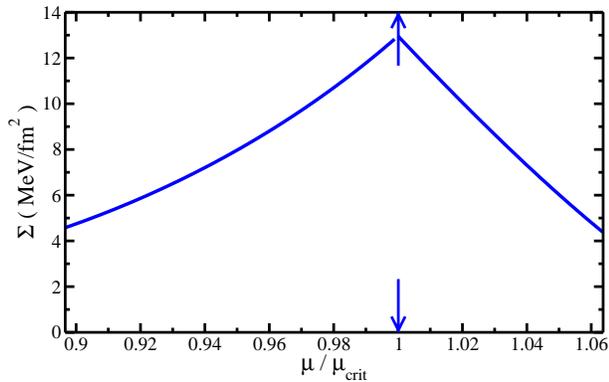}
\end{center}
\caption{Surface tension as a function of the quark chemical potential in the two metastable regions.}
\label{Sigma-fig}
\end{figure}

The surface tension is a key parameter in quantifying the nucleation process in a given medium
since it measures the amount of energy per unit of area that is spent in the construction of
a surface between the phases. In Fig. \ref{Sigma-fig}
we show our result for the surface tension, $\Sigma$,
for the chiral transition in the cold and dense LSM with quarks. The surface tension assumes values between 
$\sim 4$ and $\sim 13~$ MeV/fm$^2$, being, throughout the whole metastable region, of the order of magnitude that renders the formation of
quark matter viable during a supernova explosion, according to Ref. \cite{Sagert:2008ka}.
The biggest values occur near criticality,
since this domain is characterized by large barriers and a small free energy difference between the true and false vacua.
It should be noted
that those values are well bellow previous estimates of this quantity, as will be discussed in the conclusion.

\vspace{0.3cm}
%%%%%%%%%%%%%%%%%%%%%%%%%%%%%
%%%%%%%%%%%%%%%%%%%%%%%%%%%%%

\begin{figure}[!hbt]
\begin{center}
\vskip 0.2 cm
\includegraphics[width=8cm,angle=0]{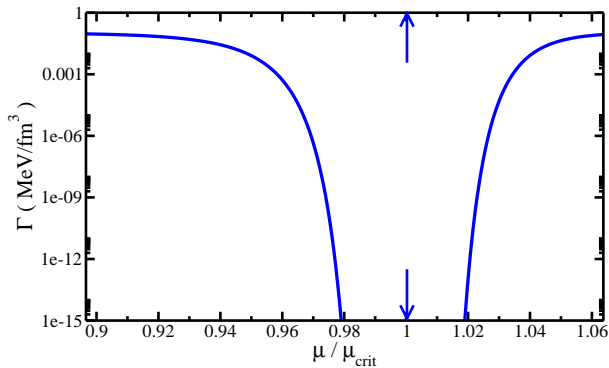}
\end{center}
\caption{Nucleation rate as a function of the quark chemical potential in the two metastable regions.}
\label{Gamma-fig}
\end{figure}

We have also estimated the nucleation rate for homogeneous nucleation \cite{langer,Csernai:1992tj} 
as $\Gamma\sim T_f^4 {\rm e}^{-F_b/T_f}$, where $F_b$ is the free energy
of the critical bubble configuration and $T_f=30~$MeV is an {\it ad hoc} temperature. Applying the expression above for the nucleation rate in this case
means that we are neglecting the temperature dependence of the critical-bubble free energy in the exponent. The results are shown in 
Fig. \ref{Gamma-fig} for both metastable regions. The nucleation rate falls abruptly as the chemical potential
approaches its critical value.

It is interesting to point out that the difference in the size of the metastable regions above and below criticality might play
an important role in determining whether nucleation is a viable process of phase conversion for a given system, as compared
to spinodal decomposition scenarios or even if the lifetime of the system represents enough time for nucleation to take place.
It is clear in Fig. \ref{Gamma-fig}, for instance, that the domain in chemical potential for which the nucleation rate
assumes sizable values is larger at the metastable region corresponding to bubble formation of hadronic matter in a partonic medium.

\vspace{0.3cm}
%%%%%%%%%%%%%%%%%%%%%%%%%%%%%
%%%%%%%%%%%%%%%%%%%%%%%%%%%%%

%%%%%%%%%%%%%%%%%%%%%%%%%%%%%%%%%%%%%%%%%%%%%%%%%%%
\subsection{Effects from vacuum terms versus thermal corrections}

Having presented in the subsection above the important parameters for nucleation and
discussed their results for the zero-temperature chiral transition in the LSMq,
let us now analyze the influence of quantum logarithmic terms in the vacuum effective potential
as well as thermal corrections.

In this section we focus on the metastable region above the critical chemical potential, aiming
for thermal nucleation of quark matter droplets in compact objects and especially supernovae explosions.
In this vein, we also keep relatively low temperatures when including thermal corrections, 
up to $T=30~$MeV. The systematic inclusion of thermal corrections provides a measure of the
validity of the zero-temperature approximation for the thermodynamics and phase structure 
in low-temperature scenarios such as those found in compact objects. We find that results for the nucleation parameters
up to $T=10~$MeV present variations within $\sim 10\%$ of the zero-temperature values, 
displayed in the previous section.

\vspace{0.3cm}
%%%%%%%%%%%%%%%%%%%%%%%%%%%%%
\begin{figure}[!h]
\begin{center}
\vskip 0.2 cm
\includegraphics[width=8cm,angle=0]{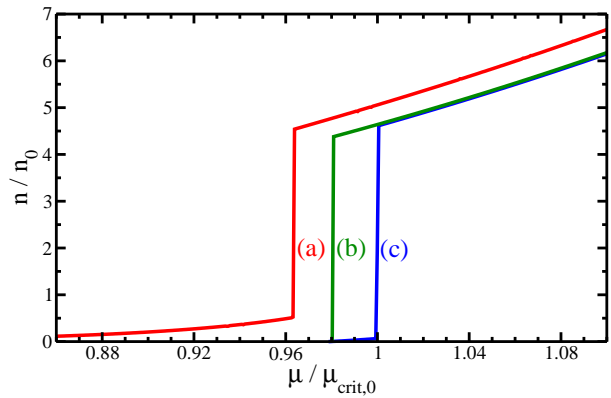}
\end{center}
\caption{Density in units of the nuclear saturation density $n_0=0.16~$fm$^{-3}$ as a function of the quark chemical potential in the cold and dense LSM (curve $(c)$), including vacuum terms (curve $(b)$) and thermal corrections (curve $(a)$).}
\label{All-Density-fig}
\end{figure}
%%%%%%%%%%%%%%%%%%%%%%%%%%%%%

\begin{figure}[!h]
\begin{center}
\vskip 0.2 cm
\includegraphics[width=8cm,angle=0]{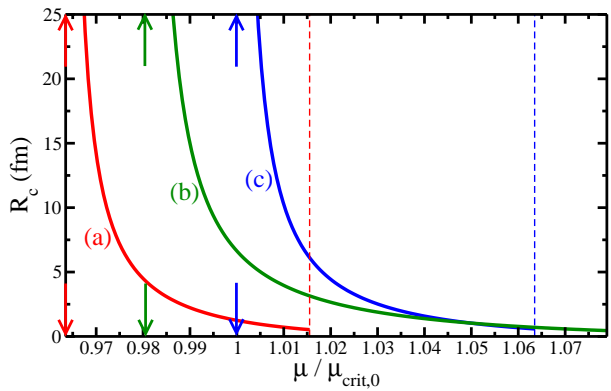}
\end{center}
\caption{Critical radius in the nucleation region of quark droplets as a function of the quark chemical potential in the cold and dense LSM (curve $(c)$), including vacuum terms (curve $(b)$) and thermal corrections (curve $(a)$).}
\label{All-Rc-fig}
\end{figure}

\begin{figure}[!h]
\begin{center}
\vskip 0.2 cm
\includegraphics[width=8cm,angle=0]{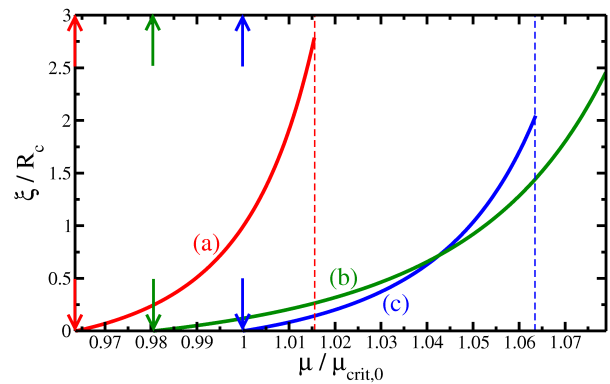}
\end{center}
\caption{Ratio between the correlation length $\xi$ and the critical radius $R_c$ as a function of the 
quark chemical potential in the cold and dense LSM (curve $(c)$), including vacuum terms (curve $(b)$) 
and thermal corrections (curve $(a)$).}
\label{All-Ratio-fig}
\end{figure}

\begin{figure}[!h]
\begin{center}
\vskip 0.2 cm
\includegraphics[width=8cm,angle=0]{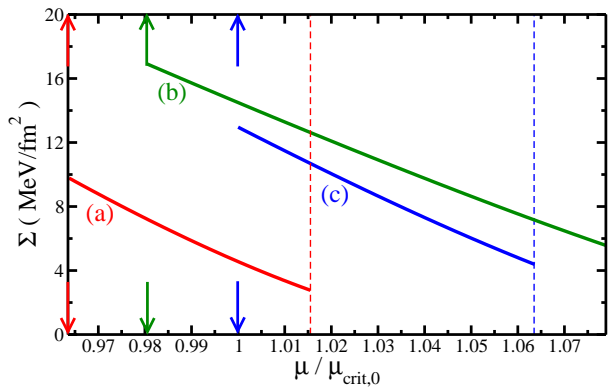}
\end{center}
\caption{Surface tension in the nucleation region of quark droplets as a function of the quark chemical potential in the cold and dense LSM (curve $(c)$), including vacuum terms (curve $(b)$) and thermal corrections (curve $(a)$).}
\label{All-Sigma-fig}
\end{figure}

\begin{figure}[!hbt]
\begin{center}
\vskip 0.2 cm
\includegraphics[width=8cm,angle=0]{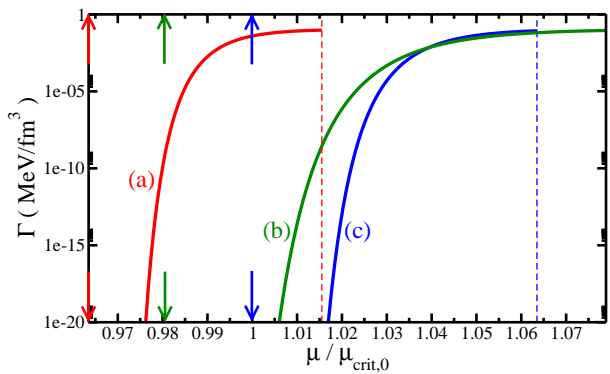}
\end{center}
\caption{Nucleation rate in the nucleation region of quark droplets as a function of the quark chemical potential in the cold and dense LSM (curve $(c)$), including vacuum terms (curve $(b)$) and thermal corrections (curve $(a)$).}
\label{All-Gamma-fig}
\end{figure}

In Figures \ref{All-Rc-fig}--\ref{All-Gamma-fig}, the role played by temperature corrections and quantum vacuum effects
is presented. More specifically, we compare results for the three following situations: $(a)$ classical vacuum
effective potential with thermal corrections for $T=30~$MeV, $(b)$ quantum vacuum effective potential
at zero temperature, $(c)$ classical vacuum effective potential at zero temperature, i.e. the
results shown in the last subsection. Each case is associated with a different critical chemical potential (indicated by arrows
in the plots; the normalization $\mu_{{\rm crit},0}=305.03~$MeV corresponds to the critical chemical potential of the case $(c)$, of the last subsection) and also
a different spinodal chemical potential (denoted by the vertical, dashed lines). 

\vspace{0.3cm}
%%%%%%%%%%%%%%%%%%%%%%%%%%%%%
%%%%%%%%%%%%%%%%%%%%%%%%%%%%%

Concerning the modification of the metastable regions, 
we show that the quantum vacuum corrections increase significantly the domain in chemical potential in which metastability persists; the metastable region
in the presence of quantum vacuum terms is $\sim 40\%$ bigger than the one associated with the classical vacuum effective potential.
This feature indicates that such corrections might generate large differences in the dynamic evolution of the system, even though 
the absolute value of the critical chemical potential itself shifts only $\sim 2\%$.

During the dynamical process of phase conversion, the density is increased (e.g. via gravitational pressure in the supernova explosion scenario)
and therefore the free energy required for including a new quark in the system also augments. 
The mapping between density and quark chemical potential within the cases we consider is plotted in Figure \ref{All-Density-fig}.
The discontinuities signal the first-order phase transition that restores chiral symmetry at high energies.
As the evolving system reaches a high enough density, it penetrates the metastable region above the critical chemical potential, 
where higher densities are associated with higher supercompression.

It is clear that the values of density predicted below the transition are irrealistically low.
This is a consequence of the fact that the LSMq fails to describe nuclear matter. Being an effetive theory constructed with the purpose
of describing the chiral symmetry features of strongly interacting matter, 
the LSMq does not predict the low-energy nuclear matter properties nor the gas-liquid transition that forms it.
However, since our ultimate interest is investigating the formation of quark matter in astrophysical processes involving ultra-compact objects,
the framework most suited is exactly one that describes the high-energy transitions of strong interactions, namely chiral restoration
and/or deconfinement. We choose therefore the LSMq and analyze the nucleation of chirally-symmetric droplets in a strongly interacting medium.

In order to interpret the results for the nucleation parameters, one should keep in mind the
variation of the metastable regions and its implication on the value of chemical potential that needs to be reached above the respective criticality,
i.e. the amount of supercompression.

Figure \ref{All-Rc-fig} shows the variation of the radius of the critical bubble,
while Figure \ref{All-Ratio-fig} displays the correlation length (or equivalently
the approximate width of the bubble wall) for the three cases.
A direct consequence of the shifting of the metastable domain is the fact that, for 
fixed chemical potential, the critical radius decreases significantly both when 
one includes quantum vacuum corrections (curve $(b)$) or in the presence of 
thermal effects (curve $(a)$). However, the amount of supercompression needed for reaching a
certain viable value of critical radius clearly increases when one includes quantum corrections
and decreases in the case with thermal corrections. Therefore,there is a nontrivial
competition between those effects: while quark matter nucleation is 
disfavored by the inclusion of quantum vacuum corrections, it is favored by thermal corrections to the nucleation
parameters.

\vspace{0.3cm}
%%%%%%%%%%%%%%%%%%%%%%%%%%%%%
%%%%%%%%%%%%%%%%%%%%%%%%%%%%%

\vspace{0.3cm}
%%%%%%%%%%%%%%%%%%%%%%%%%%%%%
%%%%%%%%%%%%%%%%%%%%%%%%%%%%%

The competition between vacuum and thermal corrections is more explicitly shown by our findings for the surface tension, 
displayed in Figure \ref{All-Sigma-fig}. As compared to the zero-temperature, classical
computation (curve $(c)$), the establishment of a interface between quark matter droplets and
the hadronic medium costs more with the incorporation of quantum corrections in the effective potential.
On the other hand, finite temperature terms tend to push the surface tension down, reaching a minimum 
of only $\sim 3~$MeV/fm$^2$, facilitating the surface formation. Therefore, if the environment
in the core of a compact object is cold enough
and if the effective theory including quantum corrections in the vacuum is the most suited for describing
the chiral transition, then the surface tension can be as high as $\sim 17.5~$MeV/fm$^2$ near criticality
rendering quark matter nucleation much slower and even an improbable phenomenon.

\vspace{0.3cm}
%%%%%%%%%%%%%%%%%%%%%%%%%%%%%
%%%%%%%%%%%%%%%%%%%%%%%%%%%%%

Analogous results for the nucleation rate $\Gamma$ can be seen in Figure \ref{All-Gamma-fig}. 
Once again, the amount of supercompression associated with a given chemical potential should 
be kept in mind. The nucleation rate falls abruptly for chemical potentials approaching the respective 
critical values. In the vicinity of the spinodals, the nucleation rate is dominated by the pre-exponential 
factor and reaches sizable values, around $\sim 0.1~$MeV/fm$^3$.

%%%%%%%%%%%%%%%%%%%%%%%%%%%%%%%%%%%%%%%%%%%%%%%%%%%%
\subsection{Consequences for quark-matter-induced supernova explosion scenario}

To contribute to the investigation of the possibility of nucleating quark matter droplets during the early
post-bounce stage of core collapse supernovae, we follow Ref. \cite{Mintz:2009ay} and define the
nucleation time as being the time it takes for the nucleation of a single critical bubble inside a volume
of $1$km$^3$, which is typical of the core of a protoneutron star, i.e.
\begin{equation}
\tau_{nucl}\equiv \left( \frac{1}{1 {\rm km}^3} \right) \frac{1}{\Gamma} \;.
\end{equation}

\vspace{0.3cm}
%%%%%%%%%%%%%%%%%%%%%%%%%%%%%
\begin{figure}[!h]
\begin{center}
\vskip 0.2 cm
\includegraphics[width=8cm,angle=0]{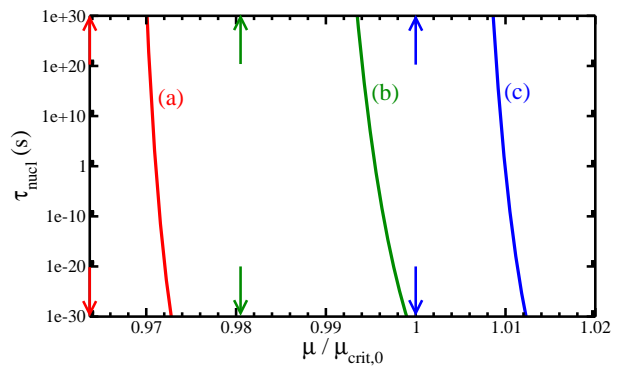}
\end{center}
\caption{Nucleation time defined for a typical volume of $1$ km$^3$ in the cold and dense LSM (curve $(c)$), including vacuum terms (curve $(b)$) and thermal corrections (curve $(a)$).}
\label{All-tau-fig}
\end{figure}
%%%%%%%%%%%%%%%%%%%%%%%%%%%%%

Fig. \ref{All-tau-fig} shows this quantity as a function of the normalized chemical potential for the three scenarios
considered in this analysis. The relevant time scale to compare is the time interval the system takes
from the critical chemical potential to the spinodal during the supernova event, if it ever reaches such
high densities in practice. Implicit in the definition above is the approximation of constant density and
temperature over the core, which is fine for an estimate, since density profiled are quite flat in this region
of the star. The typical time scale for the early post-bounce phase is of the order of a fraction of a second,
so that the time within the metastable region is smaller, in the ballpark of milliseconds.

Keeping these simplifications in mind, together with the caveat that the LSMq does not describe nuclear
matter besides the chiral properties of strong interactions, we show in Fig. 9 that vacuum corrections tend
to increase the density depth required for efficient nucleation whereas thermal corrections push in the
opposite direction, consistently with the results discussed in the previous subsection. Our results for the
nucleation time, together with those for the surface tension, tend to favor the best scenario for nucleation
of quark matter in the supernova explosion scenario considered in Ref. \cite{Mintz:2009ay}, especially when
thermal corrections with physical temperatures are included. To a great measure, this happens because
the nucleation time, and the whole process of phase conversion, depends very strongly on the surface
tension, since it enters cubed in the Boltzmann exponential of the rate $\Gamma$.

%%%%%%%%%%%%%%%%%%%%%%%%%%%%%%%%%%%%%%%%%%
\section{Conclusions and outlook}

In this paper we have computed the effective potential for the linear sigma model with 
quarks in the $\overline{\rm MS}$ scheme at zero or low temperature and 
finite quark chemical potential, including vacuum and medium fluctuations. We have
discussed the issues related to the vacuum contributions and the parameter 
fixing in the presence of these corrections. Having the full effective potential, we studied 
homogeneous nucleation in a framework that allowed for analytic calculations, given by 
a potential fit by a quartic polynomial and the thin-wall approximation. All the relevant 
quantities were computed as functions of the chemical potential (or the baryonic density), 
the key function being the surface tension. 

The value of the surface tension for the QCD phase transitions could actually play important roles in different physical phenomena. As discussed above, if this value is small enough, quark-matter formation could occur during core-collapse supernovae explosions, providing an alternative dynamics and even an observable signal of the QCD phase transition within compact objects. Moreover, a reasonably small surface tension between partonic and hadronic phases could also contribute to allow for different compact star structures including mixed phases, as argued in Ref. \cite{Kurkela:2010yk}.

Estimates for the surface tension between a quark phase and hadron matter were 
considered previously in different contexts. In a study of the minimal interface between a 
color-flavor locked phase and nuclear matter in a first order transition, the authors of 
Ref. \cite{Alford:2001zr} use dimensional analysis and obtain $\Sigma\sim 300~$MeV/fm$^{2}$ 
assuming that the transition occurs within a fermi in thickness. Taking into account the effects 
from charge screening and structured mixed phases, the authors of Ref. \cite{Voskresensky:2002hu} 
provide estimates in the range of $50-150~$MeV/fm$^{2}$ but do not exclude smaller or larger 
values. 

In this paper we computed the surface tension as a function of quark chemical 
potential (or as a function of baryon density) within the linear sigma model, isolating
the role played by quantum vacuum terms and thermal corrections. In particular, 
we show that the model predicts a surface tension of $\Sigma \sim 5$--$15~$MeV/fm$^{2}$, 
rendering nucleation of quark matter possible during the early post-bounce stage of core collapse supernovae. 
Including temperature effects and vacuum logarithmic corrections, we find a clear competition between these
features in characterizing the dynamics of the chiral phase conversion, so that if the temperature is low enough
the consistent inclusion of vacuum corrections could help preventing the nucleation of quark matter during
the early post-bounce stage of core collapse supernovae.

As discussed in Subsection \ref{Int}, effects from interactions between the sigma field and the quarks that come about at two-loop order
could in principle contribute to this competition as a third sizable modification and should be investigated
as well.

The linear sigma model, however, does not contain essential ingredients to describe nuclear 
matter, e.g. it does not reproduce features such as the saturation density and the binding energy. 
Therefore, the results obtained in this paper should be considered with caution when applied to 
compact stars or the early universe. It is an effective theory for a first-order chiral phase transition 
in cold and dense strongly interacting matter, and allows for a clean calculation of the physical 
quantities that are relevant for homogeneous nucleation in the process of phase conversion. In 
the spirit of an effective model description, our results should be viewed as estimates that indicate 
that the surface tension is reasonably low and falls with baryon density, as one increases the 
supercompression. First-principle calculations 
in QCD in this domain are probably out of reach in the near future. Therefore, estimates within 
other effective models would be very welcome.

%%%%%%%%%%%%%%%%%%%%%%%%%%%%%%%%%%%%%%%%%%
\section*{Acknowledgments} 
We thank A. Kurkela, B. W. Mintz, G. Pagliara, P. Romatschke, J. Schaffner-Bielich and A. Vuorinen for discussions. 
This work was partially supported by CAPES, CNPq, FAPERJ and FUJB/UFRJ.

% %%%%%%%%%%%%%%%%%%%%%%%%%%%%%%%%%%%%%%%%%%
% \appendix
% 
% %%%%%%%%%%%%%%%%%%%%%
% \section{Vacuum contributions to the effective potential}
% \label{vacuum-ap}
% 
% %%%%%%%%%%%%%%%%%%%%%
% \section{Fixing parameters up to two loops}
% \label{parameters-ap}

%%%%%%%%%%%%%%%%%%%%%%%%%%%%%%%%%%%%%%%%%%
%%%%%%%%%%%%%%%%%%%%%%%%%%%%%%%%%%%%%%%%%%

\end{document}